\documentclass[epj,twocolumn]{webofc}
\usepackage[varg]{txfonts}   
\woctitle{Hadron Collider Physics symposium 2012}
\usepackage{ifthen} 
\newboolean{uprightparticles}

\def\ux85 {\mbox{UX85}\xspace}

\ifthenelse{\boolean{uprightparticles}}%
{

 \def\Ppi         {\ensuremath{\uppi}\xspace}

 \def\PDelta      {\ensuremath{\Delta}\xspace}                 
 \def\PXi      {\ensuremath{\Xi}\xspace}                 
 \def\PLambda      {\ensuremath{\Lambda}\xspace}                 
 \def\PSigma      {\ensuremath{\Sigma}\xspace}                 
 \def\POmega      {\ensuremath{\Omega}\xspace}                 
 \def\PUpsilon      {\ensuremath{\Upsilon}\xspace}                 
 
 \def\PB      {\ensuremath{\mathrm{B}}\xspace}                 
                  
 \def\PD      {\ensuremath{\mathrm{D}}\xspace}

 \def\PK      {\ensuremath{\mathrm{K}}\xspace}

 \def\Pb      {\ensuremath{\mathrm{b}}\xspace}

 \def\Pi      {\ensuremath{\mathrm{i}}\xspace}

 \def\Ps      {\ensuremath{\mathrm{s}}\xspace}

}
{

 \def\Ppi         {\ensuremath{\pi}\xspace}

 \mathchardef\PDelta="7101
 \mathchardef\PXi="7104
 \mathchardef\PLambda="7103
 \mathchardef\PSigma="7106
 \mathchardef\POmega="710A
 \mathchardef\PUpsilon="7107
                  
 \def\PB      {\ensuremath{B}\xspace}                 
                  
 \def\PD      {\ensuremath{D}\xspace}

 \def\PK      {\ensuremath{K}\xspace}

 \def\Pb      {\ensuremath{b}\xspace}

 \def\Pi      {\ensuremath{i}\xspace}

 \def\Ps      {\ensuremath{s}\xspace}

}

\def\squark    {\ensuremath{\Ps}\xspace}

\def\bquark    {\ensuremath{\Pb}\xspace}
\def\bquarkbar {\ensuremath{\overline \bquark}\xspace}
\def\bbbar     {\ensuremath{\bquark\bquarkbar}\xspace}

\def\pion  {\ensuremath{\Ppi}\xspace}

\def\pip   {\ensuremath{\pion^+}\xspace}
\def\pim   {\ensuremath{\pion^-}\xspace}

\def\pipm  {\ensuremath{\pion^\pm}\xspace}
\def\pimp  {\ensuremath{\pion^\mp}\xspace}

\def\kaon  {\ensuremath{\PK}\xspace}
  \def\Kbar  {\kern 0.2em\overline{\kern -0.2em \PK}{}\xspace}

\def\Kz    {\ensuremath{\kaon^0}\xspace}
\def\Kzb   {\ensuremath{\Kbar^0}\xspace}
\def\KzKzb {\ensuremath{\Kz \kern -0.16em \Kzb}\xspace}
\def\Kp    {\ensuremath{\kaon^+}\xspace}
\def\Km    {\ensuremath{\kaon^-}\xspace}
\def\Kpm   {\ensuremath{\kaon^\pm}\xspace}
\def\Kmp   {\ensuremath{\kaon^\mp}\xspace}
\def\KpKm  {\ensuremath{\Kp \kern -0.16em \Km}\xspace}
\def\KS    {\ensuremath{\kaon^0_{\rm\scriptscriptstyle S}}\xspace} 
 
\def\Kstarz  {\ensuremath{\kaon^{*0}}\xspace}
\def\Kstarzb {\ensuremath{\Kbar^{*0}}\xspace}

  \def\Dbar    {\kern 0.2em\overline{\kern -0.2em \PD}{}\xspace}
\def\D       {\ensuremath{\PD}\xspace}

\def\Dz      {\ensuremath{\D^0}\xspace}
\def\Dzb     {\ensuremath{\Dbar^0}\xspace}
\def\DzDzb   {\ensuremath{\Dz {\kern -0.16em \Dzb}}\xspace}
\def\Dp      {\ensuremath{\D^+}\xspace}
\def\Dm      {\ensuremath{\D^-}\xspace}

\def\DpDm    {\ensuremath{\Dp {\kern -0.16em \Dm}}\xspace}

\def\Dsp     {\ensuremath{\D^+_\squark}\xspace}
\def\Dsm     {\ensuremath{\D^-_\squark}\xspace}

\def\Dsmp    {\ensuremath{\D^{\mp}_\squark}\xspace}

\def\B       {\ensuremath{\PB}\xspace}
  \def\Bbar    {\kern 0.18em\overline{\kern -0.18em \PB}{}\xspace}

\def\Bz      {\ensuremath{\B^0}\xspace}

\def\Bu      {\ensuremath{\B^+}\xspace}
\def\Bub     {\ensuremath{\B^-}\xspace}
\def\Bp      {\ensuremath{\Bu}\xspace}
\def\Bm      {\ensuremath{\Bub}\xspace}
\def\Bpm     {\ensuremath{\B^\pm}\xspace}

\def\Bs      {\ensuremath{\B^0_\squark}\xspace}
\def\Bsb     {\ensuremath{\Bbar^0_\squark}\xspace}

  \def\Y#1S{\ensuremath{\PUpsilon{(#1S)}}\xspace}

\def\Lbar {\ensuremath{\kern 0.1em\overline{\kern -0.1em\PLambda}}\xspace}

\def\to                 {\ensuremath{\rightarrow}\xspace}

\def\CP                {\ensuremath{C\!P}\xspace}

\def\AT#1     {\ensuremath{A_{\mathrm{T}}^{#1}}\xspace}           

\def\C#1      {\ensuremath{\mathcal{C}_{#1}}\xspace}                       
\def\Cp#1     {\ensuremath{\mathcal{C}_{#1}^{'}}\xspace}                    
\def\Ceff#1   {\ensuremath{\mathcal{C}_{#1}^{\mathrm{(eff)}}}\xspace}        
\def\Cpeff#1  {\ensuremath{\mathcal{C}_{#1}^{'\mathrm{(eff)}}}\xspace}       
\def\Ope#1    {\ensuremath{\mathcal{O}_{#1}}\xspace}                       
\def\Opep#1   {\ensuremath{\mathcal{O}_{#1}^{'}}\xspace}                    

\newcommand{\tev}{\ensuremath{\mathrm{\,Te\kern -0.1em V}}\xspace}
\newcommand{\gev}{\ensuremath{\mathrm{\,Ge\kern -0.1em V}}\xspace}
\newcommand{\mev}{\ensuremath{\mathrm{\,Me\kern -0.1em V}}\xspace}
\newcommand{\kev}{\ensuremath{\mathrm{\,ke\kern -0.1em V}}\xspace}
\newcommand{\ev}{\ensuremath{\mathrm{\,e\kern -0.1em V}}\xspace}
\newcommand{\gevc}{\ensuremath{{\mathrm{\,Ge\kern -0.1em V\!/}c}}\xspace}
\newcommand{\mevc}{\ensuremath{{\mathrm{\,Me\kern -0.1em V\!/}c}}\xspace}
\newcommand{\gevcc}{\ensuremath{{\mathrm{\,Ge\kern -0.1em V\!/}c^2}}\xspace}
\newcommand{\gevgevcccc}{\ensuremath{{\mathrm{\,Ge\kern -0.1em V^2\!/}c^4}}\xspace}
\newcommand{\mevcc}{\ensuremath{{\mathrm{\,Me\kern -0.1em V\!/}c^2}}\xspace}

\def\invfb   {\ensuremath{\mbox{\,fb}^{-1}}\xspace}

\def\gsim{{~\raise.15em\hbox{$>$}\kern-.85em
          \lower.35em\hbox{$\sim$}~}\xspace}
\def\lsim{{~\raise.15em\hbox{$<$}\kern-.85em
          \lower.35em\hbox{$\sim$}~}\xspace}

\def\tell1  {TELL1\xspace}
\def\ukl1   {UKL1\xspace}

\begin{document}
\title{Measurements of the CKM angle $\gamma$ in tree-dominated decays at LHCb}

\author{Stefania Ricciardi\inst{1}\fnsep\thanks{\email{stefania.ricciardi@stfc.ac.uk}} on behalf of the LHCb Collaboration}

\institute{STFC Rutherford Appleton Laboratory, Chilton, Didcot, OX11 0QX}

\abstract{%
We review the first measurements of the CKM angle $\gamma$ from LHCb. 
These measurements have been performed with $b$-hadron decays dominated by
$b\to u$ and $b\to c$ tree-level amplitudes, 
from which $\gamma$ can be determined without theoretical uncertainties.
Precision is achieved by averaging results from $B^-\to Dh^-$ ($h= K,\pi$) 
decays with $D\to h^+h^-$, $D\to \Kp\pi^-$, and $D\to \Kp\pi^-\pi^+\pi^-$, and $D\to K_{S}^0h^+h^-$. 
Prospects for these and future measurements 
of $\gamma$ using neutral $b$-hadron decays are briefly discussed.
}
\maketitle
\section{Introduction}
\label{intro}

The measurement of the CKM angle $\gamma$ [$\gamma = \mathrm{arg(-V_{ud}V_{ub}^*/V_{cd}V_{cb}^*})$]
of the Unitarity Triangle (UT) is an important goal of the LHCb physics programme.
The angle $\gamma$ plays a unique role among all CP-violating parameters because it can be 
determined using pure tree-level decays of $B$ mesons without theoretical uncertainties. 
As tree-level measurements are expected to be insensitive 
to physics beyond the Standard Model (SM), the value of $\gamma$ determined in this way provides an important SM benchmark
against which other measurements, more likely to be affected by phyics beyond the SM, can be compared. 
The power of this approach to new physics searches relies on precise measurements.
Despite results from $B$-factories and Tevatron have improved considerably the knowledge of the UT angles and sides,
$\gamma$ is still the least-well determined angle, with an experimental uncertainty from tree-level measurements
of $9\--12^{\circ}$~\cite{bib:CKMfitter, bib:UTFIT}. This precision can be significantly improved with large datasets at LHCb. 

The angle $\gamma$ can be measured in 
a theoretically clean way by exploiting the interference between $b\to u$ and $b\to c$ tree-level transitions in decays of $b$-hadrons
with a charm meson in the final state. Many $B\to DH$ decays are suitable, where $D$ indicates a $D^0$ or \Dzb and $H$ is 
the bachelor hadronic system, e.g., a kaon or a \Kstarz in the most sensitive measurements.

The first constraints on $\gamma$ at LHCb have been obtained from time-integrated measurements of charged $B$ mesons~\cite{bib:LHCBcomb}, which do not require flavour-tagging, therefore can exploit the full statistical power of the large \bbbar production cross-section in $pp$ collisions at the LHC. In these measurements $\gamma$ is determined from the interference between $\Bp\to\Dz h^+$ and $\Bp\to\Dzb h^+$ ($h = K, \pi$)\footnote{Charge-conjugation is implied throughout this paper unless otherwise stated.}, where \Dz and \Dzb decay to a common final state.  
In addition to $\gamma$, the interference depends on two $B$ hadronic parameters, which are known as $\delta_B$, 
the relative strong phase between the two $B$ decay amplitudes, and $r_B$, the relative ratio of the 
suppressed over the favoured amplitude. 
Different methods exist to determine these and other $D$ hadronic parameters from data
without theoretical uncertainties. 
Three types of $D$ decays have been studied by LHCb, corresponding to three well-established methods to determine $\gamma$ and all the unknown parameters: the GLW method~\cite{bib:GL,bib:GW}), which uses decays to \CP eigenstates ($D\to\Kp\Km$ and $D\to\pip\pim$); the ADS method~\cite{bib:ADS}), which uses decays to quasi-specific flavour eigenstates 
($D\to \Kp\pim$ and $D\to\Kp\pim\pip\pim$); and the GGSZ method~\cite{bib:GGSZ}, which uses
self-conjugate three-body final states ($D\to\KS\pip\pim$ and $D\to\KS\Kp\Km$). 

Time-integrated measurements can also be performed with neutral $b$-hadrons when the concerned decay is self-tagged. This is the case of $\Bz\to D\Kstarz$, where the charge of the kaon from the $\Kstarz\to\Kp\pim$ decay identifies the flavour of the neutral $B$ meson. Preliminary results on these decays have recently been presented by LHCb~\cite{bib:LHCBDKstar}. 

In addition, LHCb has recently performed the first time-dependent analysis of $\Bs\to \Dsmp \Kpm$ decays~\cite{bib:LHCBDsK}, which is sensitive to $\gamma - 2\beta_s$ [$\beta_s = \mathrm{arg(-V_{ts}V_{tb}^*/V_{cs}V_{cb}^*})$].

All the presented results use the 2011 dataset, corresponding to an integrated luminosity of 1~fb$^{-1}$ collected at $\sqrt{s}=7$ TeV.

\begin{figure*}[!htb]
\centering
\includegraphics[width=14cm,clip]{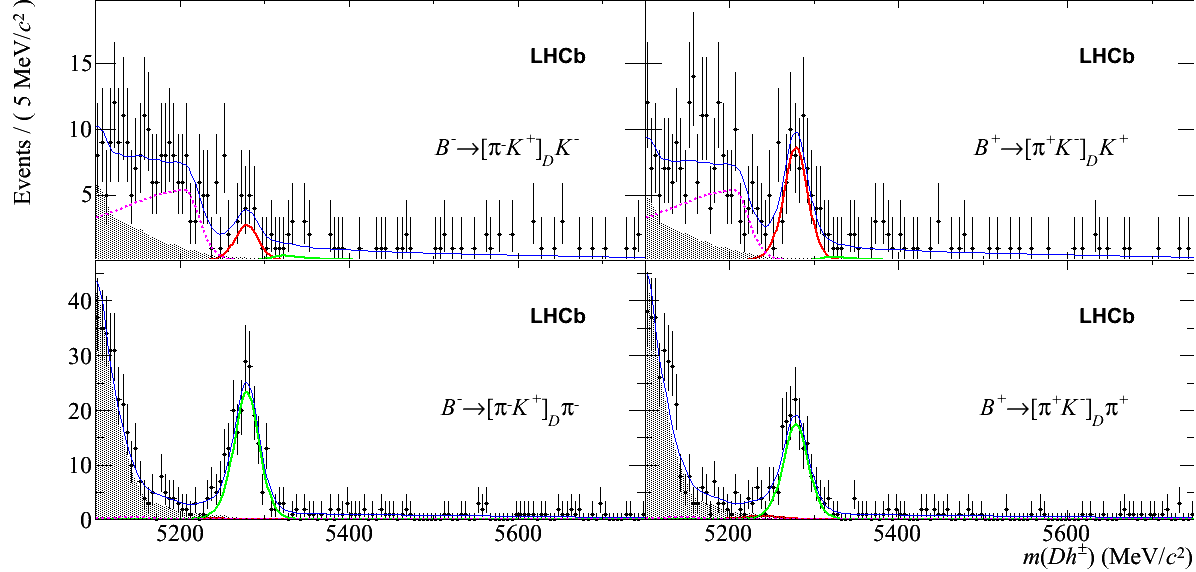}
\caption{Mass distributions of the suppressed ADS decays: (left,top) $\Bm\to D(\Kp\pim) K^-$; (right,top) $\Bp\to D(\Kp\pim) K^+$;
(left,bottom) $\Bm\to D(\Kp\pim) \pim$; and (right,bottom) $\Bp\to D(\Kp\pim) \pip$.
}
\label{fig-1}  
\end{figure*}

\section{Measurements with charged B decays}
\label{sec-1}

\subsection{ADS/GLW results} 
Charged $B$ decays to
$Dh^+$ with $D\to h^+h^-$, where each $h$ can be a kaon or a pion, are characterised by an easy topology 
and can be reconstructed with relatively high efficiency at LHCb~\cite{bib:LHCBADSGLW}. 
A combined GLW and ADS analysis is performed using the CP-even eigenstates $D\to h^+ h^-$ and 
the quasi-flavour-specific eigenstate $D\to\Kp\pim$. 
The observables with sensitivity to $\gamma$ are constructed by taking ratio of yields, so that many systematic uncertainties cancel. 
\CP asymmetries and ratios of partial widths of suppressed over favoured modes are determined with a simultaneous fit to the $B$ mass 
of all 16 possible final states. 
The expected \CP asymmetries in the $\Bp\to D \pip$ channels are smaller than the corresponding ones in 
the $\Bp\to D\Kp$ channels, since the value of $r_B$, which controls the size
of the \CP interference, is naively $\sim$20 times smaller, but 
the large yields in the $\Bp\to D \pip$ channels help constrain the mass shape in the fit.

Among the results, we note the first evidence of non-zero \CP asymmetry, $A_{\CP}$, between $\Bm\to D\Km$ and $\Bp \to D\Kp$ decays with 
$D\to \Kp\Km$, $A_{\CP} (KK) = 0.148 \pm 0.037 \pm 0.010$,
and $D\to\pip\pim$, $A_{\CP} (\pi\pi) = 0.135 \pm 0.066 \pm 0.010$, 
which combined give 4.5\,$\sigma$ significance for \CP violation in these modes.
No significant asymmetry is found in the corresponding $\Bp\to D\pip$ modes.

Another important result of this analysis is the first observation of the rare ADS mode, $\Bp\to D(\Km\pip)\Kp$,
with more than 10~$\sigma$ significance. This mode is
particularly sensitive to $\gamma$ since the two interfering amplitudes 
(i.e., the favoured $\bar{b}\to \bar{c}$  transition, followed by a doubly-Cabibbo-suppressed $D$ decay, and
the suppressed $\bar{b}\to \bar{u}$ transition, followed by the Cabibbo-favoured $D$ decay) 
have similar size, hence can give large asymmetries.
The invariant mass distribution of both the suppressed $\Bp\to D\Kp$ and $\Bp\to D\pip$ modes, 
separated by $B$ charge, are shown in Fig.~\ref{fig-1}. 
There is evidence for large \CP asymmetry in the $\Bp\to D\Kp$ mode, $A_{ADS}^K(K\pi) = -0.52 \pm 0.15 \pm 0.02$, and a hint of asymmetry in 
the $\Bp\to D\pip$, $A_{ADS}^\pi(K\pi)= 0.143 \pm 0.062 \pm 0.011$. 
Further results on GLW and ADS observables are given in Ref.~\cite{bib:LHCBADSGLW}.

More recently, LHCb has published measurements of \CP observables using the  $\Bp\to D(\Kmp\pipm\pimp\pipm)h^+$ decay~\cite{bib:LHCB4body}. 
The ADS observables share the same $B$ decay parameters ($r_B$, $\delta_B$) with $\Bp\to D(\Kmp\pipm)h^+$, but different $D$ decay parameters, 
therefore, the inclusion of this decay adds complementary information, not just additional statistics, to the determination of $\gamma$ from 
a combined fit to all the modes.
We note that the suppressed ADS modes in both the $\Bp\to D\Kp$ and $\Bp\to D\pip$ channels have not previously been observed.
In addition to the first observation of these modes at LHCb, with a significance which exceeds 5$\sigma$ and 10$\sigma$, respectively, 
we mention a hint of \CP asymmetry, $A_{ADS}^K (K3\pi) = -0.42 \pm 0.22 $ in the $\Bp\to D\Kp$ mode. The asymmetry 
in the $\Bp\to D\pip$ mode is measured to be $A_{ADS}^\pi (K3\pi) = 0.13 \pm 0.10$. 

\subsection{GGSZ results}
\begin{figure}[!htbp]
\centering
\includegraphics[width=4cm,clip]{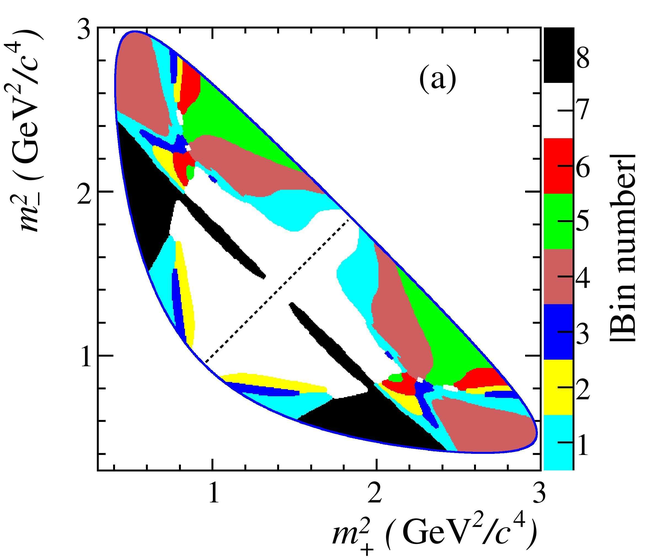}
\includegraphics[width=4cm,clip]{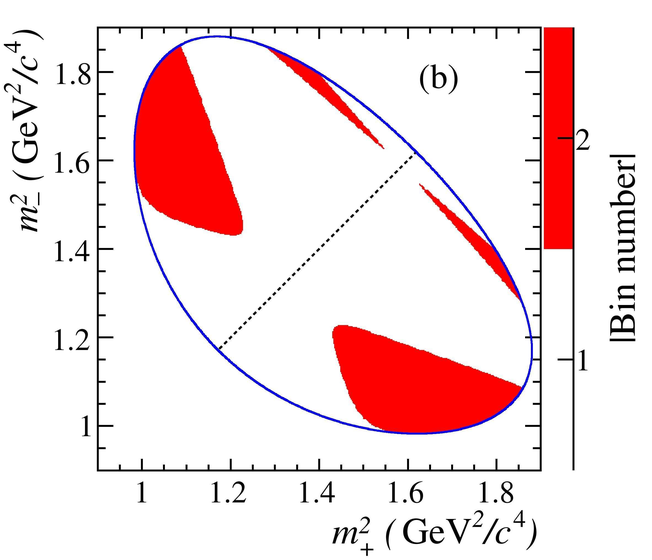}
\caption{The Dalitz plot binning shemes for  (a) $D\to\KS\pip\pim$ and (b)  $D\to\KS\Kp\Km$ decays.}
\label{fig-2}  
\end{figure}

The GGSZ method exploits the  
different interference pattern, for $\Bp\to D\Kp$ and $\Bm\to D\Km$ decays, in the $D\to\KS h^+h^-$ Dalitz plot.
This is a powerful method which dominates the sensitivity on $\gamma$ at $B$-factories, thanks to the 
rich resonance structure and the relatively large branching fraction of the $D\to \KS\pip\pim$ decay. 
The determination of $\gamma$ requires external information on the variation of the 
$D\to\KS h^+h^-$ amplitude phase over the Dalitz plot, $\delta_D$. 
A model-independent approach is taken in the first LHCb analysis with this method, which uses the CLEO~\cite{bib:CLEO3body}  
measurements of $\delta_D$ in bins of the Dalitz plot. The cartesian coordinates
$x_{\pm} = r_B \cos\,(\delta_B \pm \gamma)$ and $y_{\pm} = r_B\sin\,(\delta_B \pm \gamma)$,  are then extracted from 
a simultaneous fit to the $\Bpm$ mass distribution in each Dalitz-plot bin. 
The signal yields in each bin, $N(B^{\pm})_{i}$, are related to the cartesian coordinates by the following relations
$$ N(B^{+})_{\pm i} \propto K_{\mp i} + (x_+^2 + y_+^2)K_{\pm i} + 2\sqrt{K_{+i}K_{-i}}(x_+ c_{\pm i} \mp y_+ s_{\pm i}),$$
and
$$ N(B^{-})_{\pm i} \propto K_{\pm i} + (x_-^2 + y_-^2)K_{\mp i} + 2\sqrt{K_{+i}K_{-i}}(x_- c_{\pm i} \pm y_- s_{\pm i}),$$
where $c_i$ and $s_i$ are the amplitude weighted average cosine and sine of the strong phase difference between 
the \Dz and \Dzb decay in bin $i$, and $K_i$ is the number of events in bin $i$ of a flavour-tagged 
$\Dz\to\KS\pip\pim$ Dalitz plot. 
The index $\pm i$ varies over the number of bins, $\pm i=\pm 1,\pm 8(\pm 2)$ for $D\to\KS\pip\pim$($D\to\KS\Kp\Km$).
The chosen binning schemes are shown in Fig.~\ref{fig-2}. 
The observed $\Bp\to D\Kp$ yields are 690 for $D\to\KS\pip\pim$ and 110 $D\to\KS\Kp\Km$. 
In this analysis, the $\Bpm\to D\pipm$ modes are used not only to constrain the $\B\to DK$ mass shape in the fit, but are also used to determine the variation in the reconstruction efficiency over the Dalitz plot. To do so the assumption of no \CP violation in these decays is made and a systematic uncertainty is assigned.

\begin{figure}
\centering
\includegraphics[width=7cm,clip]{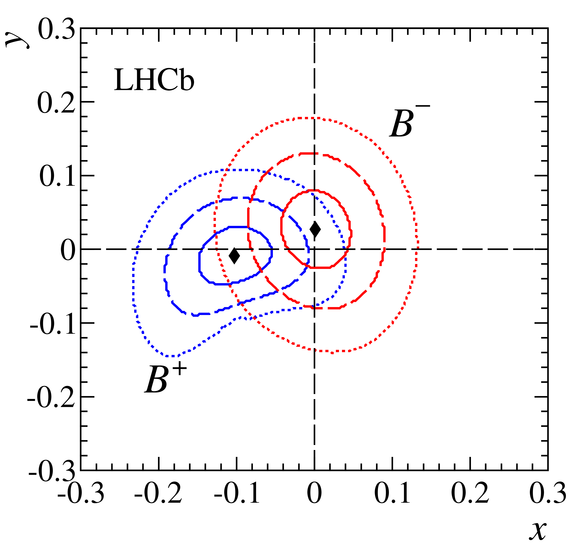}
\caption{Cartesian coordinates results for (blue) $\Bp \to D(\KS h^+h^-) \Kp$ and (red) $\Bm \to D(\KS h^+h^-) \Km$
: (solid) one, (dashed) two, and (dotted) three standard deviation confidence levels (statistical only). 
The points represent the best fit central values.
}
\label{fig-3}  
\end{figure}

\begin{figure}[hbtp!]
\centering
\includegraphics[width=7cm,clip]{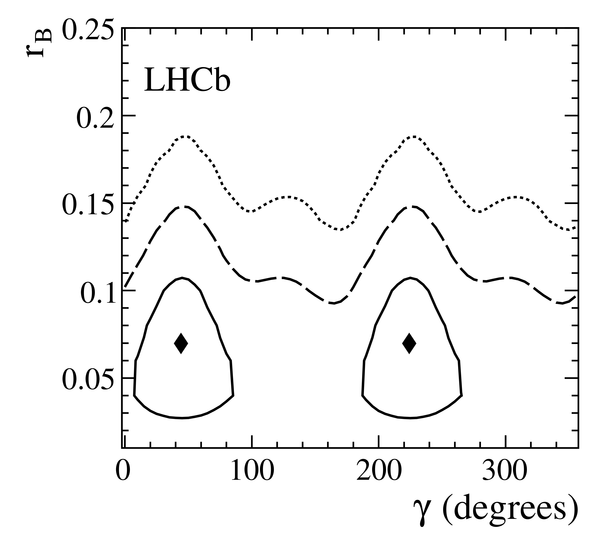}
\caption{Two-dimensional projection of confidence regions onto the ($\gamma, r_B$) plane showing the 
one (solid), two (dashed), three (dotted)  standard deviation contours from the GGSZ method alone. 
The point marks are the central values.}
\label{fig-4}       
\end{figure}

The best fit values for $x_\pm$ and $y_\pm$ are given in Ref.~\cite{bib:LHCBGGSZ} and shown in Fig.~\ref{fig-3}
together with 1, 2, 3$\,\sigma$ contours for the statistical uncertainty which are obtained from the likelihood scan. 
A non-zero value of the angle between the two vectors joining the origin in the $x$--$y$ plane and the best fit values 
is a signature of \CP violation. The current data are compatible with both the \CP violation hypothesis and its absence.
A frequentist approach is used to determine $\gamma$, $r_B$ and $\delta_B$ from the results on $x_\pm$ and $y_\pm$. 
The solution for the physics parameters has a two-fold ambiguity, ($\gamma$,  $\delta_B$) and  ($\gamma + 180^\circ$, $\delta_B + 180^\circ$).
The solution that satisfies $0 < \gamma < 180^\circ$ is $\gamma = (44^{\,+\,43}_{\,-\,38})^\circ$, $r_B = 0.07 \pm 0.04$ and $\delta_B 
= (137^{\,+\,35}_{\,-\,46})^\circ$. The value of $r_B$ is consistent with, but lower than, the world average of results 
from previous experiments~\cite{Nakamura:2010zzi}. This low value and its large correlation with the other physics parameters, also visible
in Fig.~\ref{fig-4}, explain the large uncertainty on the values of $\gamma$ obtained from this analysis. 
More stringent contraints are obtained when these results are combined with the GLW and ADS measurements, which 
have complementary sensitivity to the same physics parameters.

\section{The LHCb $\gamma$ average}
\label{sec-2}

Precision on $\gamma$ is achieved by combining all the results obtained with charged $B$ decays that are mentioned in Sect.~\ref{sec-1}.
A frequentist approach has been used by LHCb~\cite{bib:LHCBcomb}. 
The combination uses additional inputs from CLEO~\cite{bib:CLEO4body} 
for the $D\to \Kp\pim$ and $D\to\Kp\pip\pim\pip$ decay parameters. 
The recent evidence for a difference in the \CP asymmetries in $\D\to\Kp\Km$ and $D\to\pip\pim$,
$\Delta\,a_{CP}^{dir} = (-0.656 \pm 0.154) \times 10^{-2}$~\cite{bib:HFAG},  
is taken into account in the combination, however it has only marginal effects on the final results. 

\begin{figure}[hbtp!]
\centering
\includegraphics[width=6cm,clip]{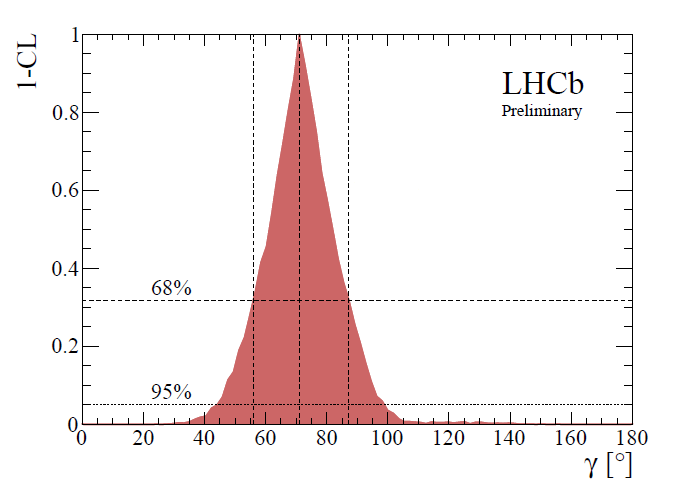}
\caption{1-C.L. curve for $\gamma$ for the combination of the measurements using $\Bp\to D\Kp$ decays.
}
\label{fig-5}       
\end{figure}

\begin{figure}[hbtp!]
\centering
\includegraphics[width=6cm,clip]{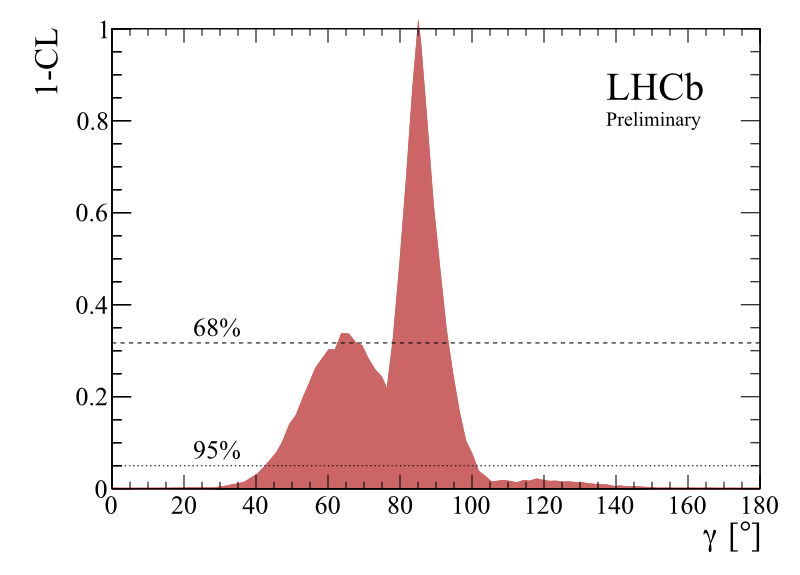}
\caption{
1-C.L. curve for $\gamma$ for the combination of the measurements using both $\Bp\to D\Kp$ and $\Bp\to D\pip$ decays.}
\label{fig-6}       
\end{figure}

The combination of all the LHCb results from $\Bp\to D\Kp$ decays gives at 68\% C.L.
$$\gamma = 71.1^{\,+\,16.6}_{\,-\,15.7} (^\circ),$$
$$ r_B = 0.092 \pm 0.008,$$
$$\delta_B = 112.0^{\,+\,12.6}_{\,-\,15.5} (^\circ).$$
Both the central value of $\gamma$  and its uncertainty are 
in good agreement with the averages recently published by BaBar~\cite{bib:BaBarAverage} 
and Belle~\cite{bib:BelleAverage}.
While results from $\Bp\to D\Kp$ exhibit approximately Gaussian behaviour, as shown in Fig,~\ref{fig-5}, 
a double-peaked structure in the $\gamma$ 1-C.L. plot arises when the GLW/ADS results 
for $\Bp\to D\pip$ are included in the combination, 
as shown in Fig.~\ref{fig-6}. 
We set the confidence limits of
$$\gamma \in [61.8, 67.8]^\circ \mathrm{ or\ } [77.9,92.4]^\circ {\mathrm{at\ 68\%\ C.L.}},$$ 
$$\gamma \in [43.8, 101.5]^\circ {\mathrm{at\ 95\%\ C.L.}},$$
where all values are modulo $180^\circ$. 
We note that the best fit value
shifts to $85^\circ$ but the 95\% confidence level is essentially unchanged.
Despite the combined value is less than $~1\,\sigma$ from the $\Bp\to D\Kp$ best fit value,  
the impact of $\Bp\to D\pip$ channel in the combination is probably larger than naively expected, 
and corresponds to rather high values of $r_B (D\pip)$ ($\in$ [0.010--0.024] at 64\% C.L.), which are preferred by the data. 
LHCb is the only experiment which has included $\Bp\to D\pip$ results in the $\gamma$ average.
Understanding and eventually fully exploiting 
$\Bp\to D\pip$ decays will be investigated with the analysis of the 2012 dataset.

\begin{figure}[hbpt!]
\centering
\includegraphics[width=7.5cm,clip]{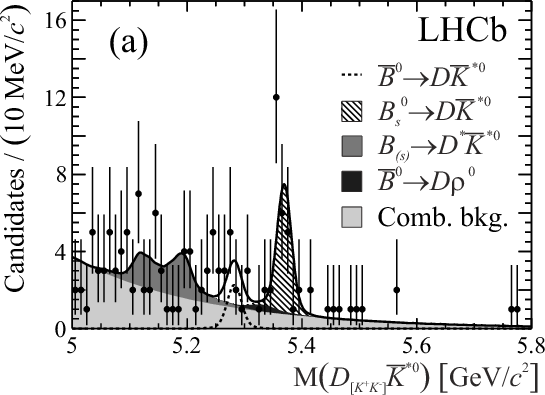}
\includegraphics[width=7.5cm,clip]{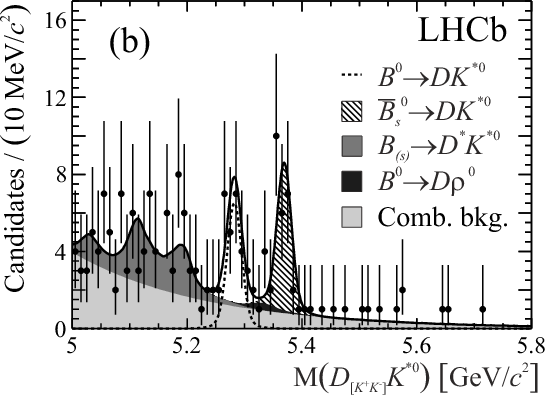}
\caption{Invariant mass distributions of (a) $D_{[\Kp\Km]}\Kstarzb$ and (b) $D_{[\Kp\Km]}\Kstarz$ candidates. The fit functions are superimposed.
}
\label{fig-7}       
\end{figure}

\section{Measurements with $B^0$ decays}
\label{sec-3}

The use of $\Bz\to D\Kstarz$ decays is particularly interesting because $r_B$ for this mode is naively a factor $\sim$ 3 larger than the analogous ratio for $\Bp\to D\Kp$ decays, hence the system can exhibit large \CP-violating effects.
The invariant mass distribution of the selected $D(K^+K^-)\Kstarz$ candidates is shown in Fig.~\ref{fig-7}. 
In addition to the \Bz mass peak, also a peak at the \Bs mass is observed.
The $\Bsb\to D\Kstarz$ decay is used as control channel, as no large \CP violation is expected in this mode.
The \CP asymmetries, computed from the efficiency-corrected signal yields, are found to be
$$A_{CP}^d = -0.45 \pm 0.23 \pm 0.02,$$
$$A_{CP}^s = 0.04  \pm 0.16 \pm 0.01,$$
for \Bz and \Bsb, respectively.
Other results for the GLW observables can be found in Ref.~\cite{bib:LHCBDKstar}. 
These are the first measurements of \CP asymmetries in \Bz and \Bsb to $D\Kstarz$ decays with the neutral $D$ meson 
decaying into a \CP-even final state. With more data, improved measurements of these and other quantities in 
$\Bz\to D\Kstarz$ decays will set important constraints on $\gamma$.

\section{Time-dependent measurements in the \Bs system}
\label{sec-4}

Interference effects are expected to be large also in the tree-level decays $\Bs\to \Dsmp\Kpm$, since the $\Bs\to \Dsmp\Kpm$
and $\Bsb\to \Dsmp\Kpm$ interfering amplitudes are of the same order in the Wolfenstein parameter $\lambda$, ${\cal {O}}(\lambda^3)$.
A time-dependent analysis is performed as the two final states of interest 
are accessible by both \Bs and \Bsb. The decay time-evolution is sensitive to $\gamma -2\beta_s$,
where $2\beta_s$ is the \Bs mixing phase, which is small in the SM and is measured with 
$\Bs\to J\psi\phi$ decays. 
 
\begin{figure}[htb!]
\centering
\includegraphics[width=8cm,clip]{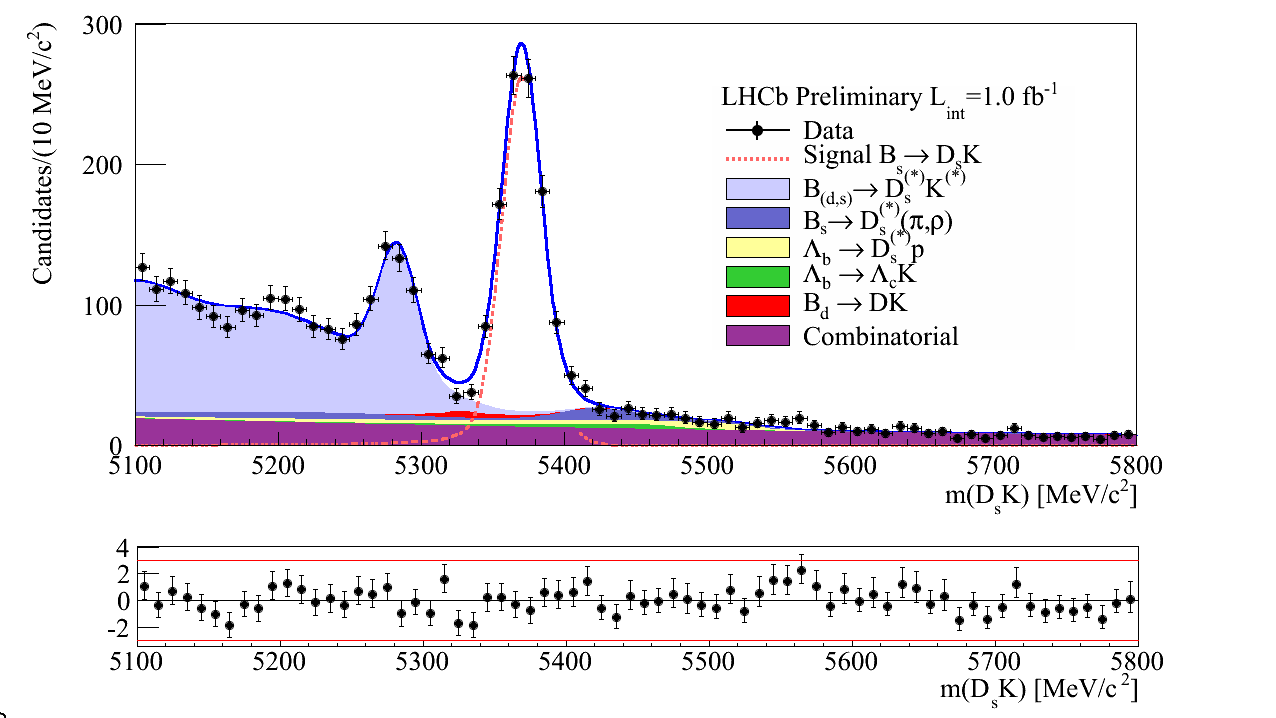}
\caption{Results of the mass fit to $\Bs\to\Dsp\Km$ candidates.
}
\label{fig-8} 
\end{figure}

\begin{figure}[htb!]
\centering
\includegraphics[width=8cm,clip]{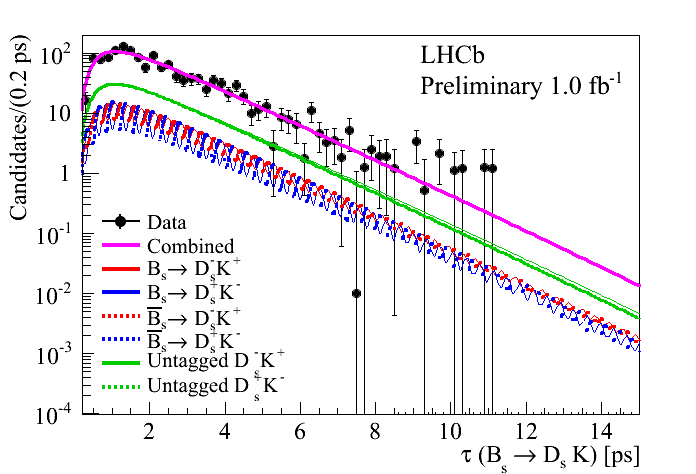}
\caption{Results of the proper-time fit to $\Bs\to\Dsp\Km$ candidates. The top magenta curve represents the result of the combined fit to 
all the samples. The green solid (dashed) curve represents the result for the untagged $\Dsm\Kp$ ($\Dsp\Km$) sample. 
The four flavour-tagged samples are represented with red and blue curves corresponding to $\Dsm\Kp$ and $\Dsp\Km$, respectively; solid (dashed) lines are used for candidates tagged as \Bs (\Bsb) at production.
}
\label{fig-9} 
\end{figure}
The first preliminary measurement of time-dependent \CP violation observables in $\Bs\to\Dsmp\Kpm$ decays 
has recently been reported by LHCb~\cite{bib:LHCBDsK}. This measurement is unique to LHCb, 
because it is the only existing experiment which 
has accumulated a sufficiently large sample of \Bs decays and is able  
to distinguish the rapid $\Bs$--$\Bsb$ oscillations (thanks to its excellent proper-time resolution and 
the boost of the \Bs mesons at LHC). 
Additional experimental challenges for this measurement are related 
to the efficiency of the flavour-tagging algorithms and the determination of the proper-time acceptance.
The LHCb results include both untagged and flavour-tagged candidates. 
Untagged events carry additional sensitivity to the weak phase since
the decay width difference in the \Bs system, $\Delta\Gamma_s$, is sizeably different from zero.
The mass distribution of the total
sample is shown in Fig.~\ref{fig-8}, and the proper-time distribution in Fig.~\ref{fig-9}, where 
the contribution of all the different components to the combined sample can be appreciated.
The results on the \CP observables can be found in~\cite{bib:LHCBDsK}.
The correlations between the systematic uncertainties have an important impact on the determination of $\gamma - 2\beta_s$
and require further studies. No attempt has been made at this stage to derive confidence intervals for the weak phase.

\section{Conclusions and prospects}
\label{sec-5}

Using 1~\invfb of the 2011 dataset, 
LHCb obtains $\gamma = 71.1^{\,+\,16.6}_{\,-\,15.7} (^\circ)$ by combining GLW, ADS and GGSZ 
$\Bp\to D\Kp$ observables. This result is in good agreement and has comparable 
precision to the combined results recently published by BaBar~\cite{bib:BaBarAverage} and Belle~\cite{bib:BelleAverage}.
Time-integrated measurements with $B^0\to D\Kstarz$ and time-dependent measurement with $\Bs\to \D_s^\pm\Kmp$ 
decays are also showing promising results. 
Statistical uncertainties are the dominant uncertainties in all these measurements. 
Precision on $\gamma$ is expected to improve significantly 
in the next years using these and other tree-level decays, 
which will become accessible with larger datasets.
It has been estimated that sub-degree precision can
be achieved with 50~\invfb at the proposed LHCb upgrade
using a combination of several decay channels~\cite{bib:LHCBimplications}.
 
\section*{Acknowledgements}
I would like to thank the organisers of HCP2012 for the excellent conference and warm hospitality in beautiful Kyoto.
\addcontentsline{toc}{section}{References}
\bibliography{main}

\begin{thebibliography}{19}

\bibitem{bib:CKMfitter}
J.~Charles et~al. (CKMfitter Group), Eur.Phys.J. \textbf{C41}, 1 (2005),
  updated results and plots available at: http://ckmfitter.in2p3.fr

\bibitem{bib:UTFIT}
D.~Derkach (UTfit Collaboration), arXiv:1301.3300  (2013)

\bibitem{bib:LHCBcomb}
R.~Aaij et~al. (LHCb collaboration), LHCb-CONF-2012-032  (2012)

\bibitem{bib:GL}
M.~Gronau, D.~London, Phys.Lett. \textbf{B253}, 483 (1991)

\bibitem{bib:GW}
M.~Gronau, D.~Wyler, Phys.Lett. \textbf{B265}, 172 (1991)

\bibitem{bib:ADS}
D.~Atwood, I.~Dunietz, A.~Soni, Phys.Rev.Lett. \textbf{78}, 3257 (1997)

\bibitem{bib:GGSZ}
A.~Giri, Y.~Grossman, A.~Soffer, J.~Zupan, Phys.Rev. \textbf{D68}, 054018
  (2003)

\bibitem{bib:LHCBDKstar}
R.~Aaij et~al. (LHCb collaboration) (2012), \texttt{arXiv:1212.5205}

\bibitem{bib:LHCBDsK}
R.~Aaij et~al. (LHCb collaboration), LHCb-CONF-2012-029  (2012)

\bibitem{bib:LHCBADSGLW}
R.~Aaij et~al. (LHCb Collaboration), Phys.Lett. \textbf{B712}, 203 (2012),
  erratum Phys. Lett. \textbf{B713} (2012) 351

\bibitem{bib:LHCB4body}
R.~Aaij et~al. (LHCb collaboration), LHCb-CONF-2012-030  (2012)

\bibitem{bib:CLEO3body}
J.~Libby et~al. (CLEO Collaboration), Phys.Rev. \textbf{D82}, 112006 (2010)

\bibitem{bib:LHCBGGSZ}
R.~Aaij et~al., Phys. Lett. \textbf{B718}, 43 (2012)

\bibitem{Nakamura:2010zzi}
J.~Beringer et~al. (Particle Data Group), Phys.Rev. \textbf{D86}, 010001 (2012)

\bibitem{bib:CLEO4body}
N.~Lowrey et~al. (CLEO Collaboration), Phys.Rev. \textbf{D80}, 031105 (2009)

\bibitem{bib:HFAG}
Y.~Amhis et~al. (Heavy Flavor Averaging Group) (2012), \texttt{arXiv:1207.1158}

\bibitem{bib:BaBarAverage}
J.~Lees et~al. (BABAR Collaboration) (2013), \texttt{arXiv:1301.1029}

\bibitem{bib:BelleAverage}
K.~Trabelsi (Belle collaboration) (2013), \texttt{arXiv:1301.2033}

\bibitem{bib:LHCBimplications}
R.~Aaij et~al. (LHCb Collaboration) (2012), \texttt{arXiv:1208.3355}

\end{thebibliography}
\end{document}